\documentclass[a4paper,11pt]{article}
\usepackage{latexsym}
\usepackage{amssymb}
\usepackage{amsmath}
\usepackage{cite}
\usepackage{authblk}
\usepackage{graphicx}
\usepackage{subcaption}
\usepackage{caption,subcaption}
\usepackage{epsfig}
\usepackage{float}
\usepackage{subcaption}
\title{Cosmic acceleration in screening hybrid quintessence model}

\author[1]{H. Mohseni Sadjadi\thanks{mohsenisad@ut.ac.ir}}

\affil[1]{Department of Physics, College of Science, University of Tehran, P.O.B. 14395/547, Tehran, Iran}

\begin{document}
\maketitle

\begin{abstract}
A formalism for the appearance of dark energy in the matter dominated era, leading to a persistent de Sitter expansion at the late time is proposed. Our framework is a hybrid quintessence model with a nonminimal coupling to the Ricci scalar. Coupling to the curvature leads to the screening effect and triggers the dark energy evolution. The coupling of the scalar fields drives this evolution to a de Sitter stable fixed point. These occur via successive $Z_2$ symmetry breakings.
\end{abstract}
\section{Introduction}
Astrophysical data indicate that the expansion of the Universe is accelerating \cite{acc1,acc2}.  The origin of this positive acceleration which has begun from the matter dominated era is still unclear but may be caused by a scalar field, whose pressure is negative, called quintessence \cite{quint1,quint2,quint3,quint4,quint5,quint6,quint7,quint8,quint9,quint10,quint11,quint12}.

The negativity of the pressure implies that the dark energy density redshifts slower than other densities including dark matter. But dark matter and dark energy densities have the same order of magnitude today \cite{coin1,coin2,coin3,coin4,coin5,coin6,coin7,coin8}. Therefore, in the early era, the dark energy density was negligible. If we consider a dynamical dark energy, a question arises: why the quintessence has become relevant in the recent era? What led to the emergence of dark energy from a matter dominated epoch?  To find an answer to this question, one might be inspired by the screening models.  These models explain how a field, $\phi$, is inactive in a dense environment while active in a dilute medium. In this view, dilution of matter due to the Universe expansion, stimulated the quintessence evolution.  One of the screening models is the symmetron model in which the quintessence interacts with matter via a conformal coupling, $\tilde{g}_{\mu \nu}=A(\phi^2)g_{\mu \nu}$ ($A$ is a positive real analytical function). If the matter density becomes less than a critical value, $Z_2$ symmetry breaks and the field $\phi$ becomes active.
There are some problems in applying this model to describe the recent cosmic acceleration:  The quintessence after a while rolls down towards the minimum of the potential and by oscillating about it ceases to behave as dark energy.  In addition, the positive acceleration requires a positive potential, but the symmetry breaking reduces the potential, so we need to add a positive term to the potential playing the role of a cosmological constant \cite{sym,khou,sam,hms1,hms2}, which makes the problem similar to $\Lambda$CDM, reducing the motive of using the scalar field.  Recently, this model was modified by including a nonminimal derivative coupling term in the kinetic part of the action, which slows down the quintessence and prolongs the acceleration period \cite{mat}. In this model too, the positive acceleration is not persistent, and considering a cosmological constant term, to drive the cosmic acceleration, is necessary.
In \cite{sad1,sad2}, the symmetron with a nonminimal coupling to the scalar torsion $(\propto T\phi^2)$ has been considered in the context of the teleparallel model of gravity. The coupling to the torsion dramatically changes the quintessence equation of motion and also the Friedmann equations, enabling the model to have a de Sitter expansion at the late time. This attractor de Sitter solution is absent in the standard symmetron model. In addition, there is no more need to add a cosmological constant to the problem, so our goal to consider a dynamical dark energy is not messed up. In this model too, the dark energy density is ignorable in the early matter dominated era and the quintessence evolution is ignited by the conformal coupling which breaks the $Z_2$ symmetry breaking. In this model, due to the nonminimal coupling of the scalar field to the torsion scalar, the Lorentz invariance is violated. It may seem that if instead of the torsion scalar we were using the Ricci scalar, the Lorentz invariance would not be destroyed and we had a better model. Such a quintessence-Ricci scalar coupling $(\propto R\phi^2)$ has also been introduced before in the context of renormalizability of quantum scalar field models \cite{cal}. In this context, the symmetron model was discussed in \cite{sad1}, and there was shown that the model cannot have a stable attractor solution like that has appeared in the teleparallel model. So the symmetron model in this framework also leads to a short period of acceleration \cite{hyb1}.

In this paper we try to get a mechanism based on the symmetry breaking, to explain the Universe acceleration in a Lorentz invariant framework.  We consider a coupling between the scalar field and the Ricci scalar, which plays the role of an additional mass term for the quintessence. The matter density is imprinted in the Ricci scalar, so to the occurrence of a symmetry breaking due to the matter density dilution, does not require any additional conformal coupling. Although the nonminimal coupling can explain the onset of dark energy, the acceleration, like the symmetron model, is not persistent. To obtain a stable de Sitter expansion for the late time, we employ an additional scalar field and construct a hybrid model. The history of employing the hybrid model goes back to the hybrid inflation \cite{infl}, in which, the inflation ends due to the rapid rolling (waterfall) of a scalar field triggered by another scalar field. Hybrid models have also been employed in the context of dark energy, to describe the late time acceleration \cite{hybdark} and also to describe the possible phantom divide line crossing \cite{hybcr}.
In our model, the evolution of the dark energy is governed by two successive symmetry breakings. The first one increases the dark energy density in the matter dominated era causing the Universe to (super-) accelerate, and the second one determines the Universe evolution path at the late time, for example, it may lead the acceleration to a stable de Sitter expansion. In \cite{sad1}, this r\^{o}le was assigned to the additional nonminimal coupling to the torsion scalar which violates the Lorentz invariance. In our framework (unlike the symmetron model) there is no need to insert an additional cosmological constant term and unlike the teleparallel model \cite{sad2}, the lorentz invariance is respected.

 The paper is organized as follows: In the next section, we introduce the model and study the cosmic evolution from matter dominated era until the late time de Sitter Universe. We discuss the stability of the model. In the last section, we conclude and discuss our results and elucidate our model through a numerical example.

We use units $\hbar=c=1$.

\section{Hybrid quintessence with nonminimal coupling}
We start with the action
\begin{equation}\label{1}
S=\int\left(\frac{1}{2}(M_P^2-\epsilon\phi^2)R-\frac{1}{2}\partial_\mu \phi \partial^\mu \phi-V(\phi,\psi)-\frac{1}{2}\partial_\mu \psi \partial^\mu \psi\right)\sqrt{-g} d^4x+S_m,
\end{equation}
where $\phi$ and $\psi$ are two real scalar fields. $M_p$ is the reduced Planck mass. $\epsilon$ is a positive constant which by coupling $R$ to $\phi$, relates the dynamics of $\phi$ to non relativistic matter density contributing in $R$. Such a coupling has been employed to study the inflation as well as the late time acceleration \cite{curv1,curv2,curv3}. The scalar field $\phi$ is equally coupled to all matter components through the trace of their energy momentum tensor (note that $R=-\frac{T^\mu_\mu}{M_P^2}$, hence the model respects the weak equivalence). To study the cosmological aspects of the model, we consider (Friedmann-Lema\^{i}tre-Robertson-Walker) FLRW space-time
\begin{equation}\label{2}
ds^2=-dt^2+a^2(t)(dx^2+dy^2+dz^2).
\end{equation}
In this Universe, the Ricci scalar is given by
\begin{equation}\label{3}
R=6\dot{H}+12H^2.
\end{equation}
Variation of the action with respect to the metric gives the Friedmann equations
\begin{equation}\label{4}
H^2=\frac{1}{3M_p^2}\left(\rho_{d}^{eff.}+\rho_m \right),
\end{equation}
and
\begin{equation}\label{5}
\dot{H}=-\frac{1}{2M_P^2}(P^{eff.}_d+\rho^{eff.}_d+P_m+\rho_m),
\end{equation}
where, the effective dark energy density, $\rho_{d}^{eff.}$, and the effective pressure, $P^{eff.}_d$, are given by
\begin{equation}\label{6}
\rho_{d}^{eff.}=\frac{1}{2} \dot{\phi}^2+V(\phi,\psi)+6\epsilon H \phi \dot{\phi}+3\epsilon H^2\phi^2+\frac{1}{2}\dot{\psi}^2,
\end{equation}
and
\begin{eqnarray}\label{7}
P^{eff.}_d&=&\frac{1}{2}\dot{\phi}^2+\frac{1}{2}\dot{\psi}^2-V(\phi,\psi)-2\epsilon\dot{\phi}^2-2\epsilon\dot{H}\phi^2+2\epsilon H\phi \dot{\phi}\nonumber \\
&-&(3\epsilon H^2-2\epsilon^2R)\phi^2+2\epsilon \phi V_{,\phi}
\end{eqnarray}
respectively. Variation of the action with respect to the scalar field $\phi$ gives
\begin{equation}\label{8}
\ddot{\phi}+3H\dot{\phi}+\epsilon R \phi+V_{,\phi}=0.
\end{equation}
Similarly for $\psi$ we obtain
\begin{equation}\label{9}
\ddot{\psi}+3H\dot{\psi}+V_{,\psi}=0.
\end{equation}
We assume that the matter component is constituted of  barotropic fluids $\rho_m=\sum_i\rho^{(i)}$. The continuity equations are
\begin{equation}\label{10}
\dot{\rho^{(i)}}+3H(\rho^{(i)}+P^{(i)})=0.
\end{equation}
 Hence, in terms of the scale parameter $a(t)$, the redshift of $ith$ matter component can be derived from (\ref{10}) as $\rho^{(i)}(t)=\rho_0^{(i)}a^{-3(1+w^{(i)})}$.  $w^{(i)}$ is the corresponding equation of state (EoS) parameter, and $\rho_0^{(i)}$ denotes the present density evaluated at $a=1$.
As the base of our model is the $Z_2$ symmetry breaking, we chose the potential as
\begin{equation}\label{11}
V(\phi,\psi)=-\frac{1}{2}\mu^2\phi^2+\frac{\lambda}{4}\phi^4-\frac{1}{2}\gamma \phi^2 \psi^2+\frac{1}{2}m^2\psi^2+\frac{\Lambda}{4}\psi^4.
\end{equation}

We rescale $H$  using a Hubble parameter $H_0$ ($H_0$ may be taken as the present Hubble parameter) as $\tilde {H}=\frac{H}{H_0}$,  and use dimensionless parameters $\tilde{\phi}=\frac{\phi}{M_p},\,\ \tilde{\psi}=\frac{\psi}{M_P},\,\,\tilde{\rho}=\frac{\rho}{M_P^2H_0^2},\,\, \tilde{\lambda}=\frac{\lambda M_P^2}{H_0^2},\,\, \tilde{\Lambda}=\frac{\Lambda M_P^2}{H_0^2},\,\, \tilde{\mu}=\frac{\mu}{H_0},\,\, \tilde{m}=\frac{m}{H_0},\,\, \tilde{\gamma}=\frac{\gamma M_P^2}{H_0^2},\,\,\tilde{R}=\frac{R}{H_0^2}$. In terms of these parameters and the dimensionless time $\tau=tH_0$ we have
\begin{equation}\label{12}
\tilde{V}=\frac{V}{H_0^2M_P^2}=-\frac{1}{2}\tilde{\mu}^2\tilde{\phi}^2+\frac{\tilde{\lambda}}{4}\tilde{\phi}^4-\frac{1}{2}\tilde{\gamma} \tilde{\phi}^2 \tilde{\psi}^2+\frac{1}{2}\tilde{m}^2\tilde{\psi}^2+\frac{\tilde{\Lambda}}{4}\tilde{\psi}^4.
\end{equation}
The equations of motion are
\begin{equation}\label{13}
\frac{d^2{\tilde{\phi}}}{d\tau^2}+3\tilde{H}\frac{d{\tilde{\phi}}}{d\tau}+\left(-\tilde{\gamma}\tilde{\psi}^2+\epsilon \tilde{R} -\tilde{\mu}^2\right)\tilde{\phi}+\tilde{\lambda}\tilde{\phi}^3=0,
\end{equation}
and
\begin{equation}\label{14}
\frac{d^2{\tilde{\psi}}}{d\tau^2}+3\tilde{H}\frac{d{\tilde{\psi}}}{d\tau}+\left(-\tilde{\gamma}\tilde{\phi}^2+\tilde{m}^2\right)\tilde{\psi}+\tilde{\Lambda} \tilde{\psi}^3=0.
\end{equation}
Equipped with these equations, we perform our model as follows:
Initially, we assume that the scalar fields are settled down in the stable minimum of the potential (\ref{11}),  i.e. $\phi=0$, $\psi=0$. In this stage, the dark energy density is zero and the Ricci scalar is given by  $\tilde{R}=\tilde{\rho}^{(m)}$, where by the superscript $m$, we denote dust and pressureless dark matter.  Note that radiation satisfies $P=\frac{\rho}{3}$ , so has no contribution in $R$. We assume that initially, the dominated pressureless matter density is greater than the critical density  $\tilde{\rho}^{(m)}>\tilde{\rho}_{cr.}\equiv \frac{\tilde{\mu}^2}{\epsilon}$. The equation of motion (\ref{13}), implies that we can take an effective potential for $\tilde{\phi}$ whose behavior around the initial point is determined by
\begin{equation}\label{16}
V^{eff.}_{,\phi}=\left( \tilde{\rho}^{(m)}\epsilon -\tilde{\mu}^2\right)\tilde{\phi}+\tilde{\lambda}\tilde{\phi}^3.
\end{equation}
For $\tilde{\rho}^{(m)}>\frac{\tilde{\mu}^2}{\epsilon}$, this potential is convex. By the expansion of the Universe,  $\tilde{\rho}^{(m)}\propto a^{-3}$ becomes less than the critical value, and the point $\tilde{\phi}=0$ becomes unstable. In this era the squared effective mass of $\tilde{\phi}$ is negative $\tilde{\mu}_{eff.}^2=\tilde{\rho}^{(m)}\epsilon  -\tilde{\mu}^2<0$, and the effective potential becomes concave,  and $\tilde{\phi}$ begins its evolution.  During this evolution, the Universe may experience a period of acceleration or super acceleration. Note that if we took $\epsilon<0$, this process would not be realized. In its turn, the squared effective mass of $\tilde{\psi}$ becomes negative when $\tilde{\phi}^2>\frac{\tilde{m}^2}{\tilde{\gamma}}$, and $\tilde{\psi}$ begins its evolution from $\tilde{\psi}=0$ . A fixed point of the equations (\ref{5}), (\ref{8}), (\ref{9}), and (\ref{10}) at the late time is given by
$\frac{d\tilde{\phi}}{d\tau}=\frac{d\tilde{\psi}}{d\tau}=\frac{\tilde{H}}{d\tau}=\tilde{\rho}^{(m)}=0$. This specifies a  de Sitter solution characterized by $\{\tilde{H}=\tilde{H}_{dS}, \tilde{\phi}=\tilde{\phi}_{dS}, \tilde{\psi}=\tilde{\psi}_{dS}, \tilde{\rho}^{(m)}_{dS}=0\}$, where
\begin{eqnarray}\label{17}
\tilde{\phi}_{dS}^2=-\frac{\epsilon \tilde{m}^4+\tilde{\Lambda} \tilde{\mu}^2-\tilde{\gamma} \tilde{m}^2}{\tilde{\Lambda}\epsilon \tilde{\mu}^2-\epsilon \tilde{\gamma} \tilde{m}^2-\tilde{\Lambda} \tilde{\lambda} +\tilde{\gamma}^2}\nonumber \\
\tilde{\psi}_{dS}^2=-\frac{\epsilon \tilde{m}^2\tilde{\mu}^2+\tilde{\gamma} \tilde{\mu}^2-\tilde{\lambda} \tilde{m}^2}{\tilde{\Lambda}\epsilon \tilde{\mu}^2-\epsilon \tilde{\gamma} \tilde{m}^2-\tilde{\Lambda} \tilde{\lambda} +\tilde{\gamma}^2}\nonumber \\
\tilde{H}_{dS}^2=\frac{1}{12}\frac{\tilde{\Lambda}\tilde{\mu}^4-2\tilde{\gamma} \tilde{m}^2\tilde{\mu}^2+\tilde{\lambda} \tilde{m}^4}{\tilde{\Lambda}\epsilon \tilde{\mu}^2-\epsilon \tilde{\gamma} \tilde{m}^2-\tilde{\Lambda} \tilde{\lambda} +\tilde{\gamma}^2}.
\end{eqnarray}
The existence of the solution (\ref{17}) necessitates that the parameters satisfy
\begin{eqnarray}\label{18}
-\frac{\epsilon \tilde{m}^4+\tilde{\Lambda} \tilde{\mu}^2-\tilde{\gamma} \tilde{m}^2}{\tilde{\Lambda}\epsilon \tilde{\mu}^2-\epsilon \tilde{\gamma} \tilde{m}^2-\tilde{\Lambda} \tilde{\lambda} +\tilde{\gamma}^2}>0\nonumber \\
-\frac{\epsilon \tilde{m}^2\tilde{\mu}^2+\tilde{\gamma} \tilde{\mu}^2-\tilde{\lambda} \tilde{m}^2}{\tilde{\Lambda}\epsilon \tilde{\mu}^2-\epsilon \tilde{\gamma} \tilde{m}^2-\tilde{\Lambda} \tilde{\lambda} +\tilde{\gamma}^2}>0\nonumber \\
\frac{\tilde{\Lambda}\tilde{\mu}^4-2\tilde{\gamma} \tilde{m}^2\tilde{\mu}^2+\tilde{\lambda} \tilde{m}^4}{\tilde{\Lambda}\epsilon \tilde{\mu}^2-\epsilon \tilde{\gamma} \tilde{m}^2-\tilde{\Lambda} \tilde{\lambda} +\tilde{\gamma}^2}>0.
\end{eqnarray}
If we did not consider the second field, $\tilde{\psi}$, we would have the inconsistent solution
\begin{eqnarray}\label{19}
\tilde{\phi}_{dS}^2=-\frac{\tilde{\mu}^2}{\epsilon\tilde{\mu}^2-\tilde{\lambda}}\nonumber \\
\tilde{H}_{dS}^2=\frac{1}{12}\frac{\tilde{\mu}^4}{\epsilon \tilde{\mu}^2-\tilde{\lambda}}
\end{eqnarray}
which gives either an imaginary Hubble parameter or an imaginary scalar field.  So the presence of $\tilde{\psi}$ is necessary to have a late time de Sitter expansion.

\subsection{Dynamical phase space and stability}
In this part we study whether the fixed point (\ref{17}) is stable. We define the new parameters
$x:=\frac{1}{\sqrt{6}\tilde{H}}\frac{d\tilde{\phi}}{d\tau}$, $y:=\tilde{\phi}$, $Y:=\tilde{\psi}$, $X:=\frac{1}{\sqrt{6}\tilde{H}}\frac{d\tilde{\psi}}{d\tau}$,  $u=\frac{\sqrt{\tilde{\rho}^{(m)}}}{\sqrt{3}\tilde{H}}$.
We use the equations of motion to obtain the following autonomous system of differential equations
\begin{eqnarray}\label{20}
&&x'=E_1(x,y,X,Y):=-3x-\sqrt{6}\epsilon y\left(2+s(x,y,X,Y,u)\right)-xs(x,y,X,Y,u)\nonumber \\
&&-\frac{\sqrt{6}}{2}f(y,Y)\left(1-u^2-x^2-X^2-2\sqrt{6}\epsilon x y-\epsilon y^2\right)\nonumber \\
&&y'=E_2(x,y,X,Y):=\sqrt{6}x \nonumber \\
&&u'=E_3(x,y,X,Y):=-\frac{3}{2}u-s(x,y,X,Y,u)u\nonumber \\
&&Y'=E_4(x,y,X,Y):=\sqrt{6}X\nonumber \\
&&X'=E_5(x,y,X,Y):=-3X-s(x,y,X,Y,u)X-\frac{\sqrt{6}}{2}F(y,Y)\Big(1-u^2\nonumber \\
&&-x^2-X^2-2\sqrt{6}\epsilon xy -\epsilon y^2\Big),
\end{eqnarray}
where $f(y,Y)=\frac{V_{,y}}{V}\,\,\,\,F(y,Y)=\frac{V_{,Y}}{V}$, and prime denotes derivative with respect to $\ln a$. We have also defined
\begin{eqnarray}\label{21}
&&s:=\frac{\dot{\tilde{H}}}{\tilde{H}^2}=\frac{1}{-1+\left(-6\epsilon^2+\epsilon\right)y^2}\Big(\frac{3}{2}u^2+3(1-2\epsilon)x^2+3X^2+4\sqrt{6}\epsilon xy\nonumber \\
&&+12\epsilon^2y^2+3\epsilon y f(y,Y)(1-u^2-x^2-X^2-2\sqrt{6}xy-\epsilon y^2)\Big).
\end{eqnarray}
Two fixed point of the above autonomous system are $\{x=0,X=0,s=-\frac{3}{2}\}$ for which $u\neq 0$, and $\{x=0,X=0,u=0,s=0\}$ which is the de Sitter point specified by (\ref{17}). The first critical point corresponds to our initial point from which the system began its evolution in the matter dominated era.  The stability of the de Sitter fixed point can be examined by studying small perturbations around it:  $x=x_c+\delta x,\,\, y=y_c+\delta y,\,\, u=u_c+\delta u,\,\,\, X=X_c+\delta X,\,\, Y=Y_c+\delta Y$. After some computation we obtain
\begin{equation}\label{22}
{d\over d{\ln a}}{ \left( \begin{array}{cccc}
\delta x\\
\delta y \\
\delta u \\
\delta X  \\
\delta Y
\end{array} \right)}=\mathcal{M} \left( \begin{array}{cccc}
\delta x\\
\delta y \\
\delta u \\
\delta X \\
\delta Y
\end{array} \right),
\end{equation}
where
\begin{equation}\label{23}
\mathcal{M}=\left( \begin {array}{ccccc} E_{1,x}& E_{1,y} &E_{1,u}&E_{1,X}&E_{1,Y} \\ \noalign{\medskip} E_{2,x}& E_{2,y} &E_{2,u}&E_{2,X}&E_{2,Y}\\ \noalign{\medskip}E_{3,x}& E_{3,y} &E_{3,u}&E_{3,X}&E_{3,Y}\\ \noalign{\medskip} E_{4,x}& E_{4,y} &E_{4,u}&E_{4,X}&E_{4,Y}\\ \noalign{\medskip}E_{5,x}& E_{5,y} &E_{5,u}&E_{5,X}&E_{5,Y}\end {array} \right).
\end{equation}
which must be evaluated at the fixed point, leading to
\begin{equation}\label{24}
\mathcal{M}=\left( \begin {array}{ccccc} M_{11}& M_{12} &0&M_{14}&0 \\ \noalign{\medskip}\sqrt{6}& 0 & 0&0&0\\ \noalign{\medskip}0& 0 &-\frac{3}{2}&0&0\\ \noalign{\medskip} 0& 0 &0&0&\sqrt{6}\\ \noalign{\medskip}0&M_{52}&0&M_{54}&-3\end {array} \right).
\end{equation}
where the nonzero components, after a lengthy calculation,  are obtained as
\begin{equation}\label{25}
M_{11}=\frac{3+(144\epsilon^4-162\epsilon^3+3\epsilon^2)y_c^4+(18\epsilon^2-6\epsilon)y_c^2}{-1+6(\epsilon-\frac{1}{6})\epsilon^2y_c^4+(2\epsilon-6\epsilon^2)y_c^2},
\end{equation}
\begin{eqnarray}\label{26}
&&M_{12}=-\frac{\sqrt{6}\left((\epsilon y_c^2-1)^2f_{,y_c}+4\epsilon^2 y_c^2+4\epsilon\right)}{12\epsilon^2y_c^2-2\epsilon y_c^2+2}\nonumber \\
&&=\sqrt{6}\Big((\epsilon y_c^2-1)^2\Big(4Y_c^6\tilde{\gamma}^2\tilde{\Lambda}+((8\tilde{\gamma}^4-12\tilde{\Lambda}\tilde{\lambda})y_c^2+8\tilde{m}^2\tilde{\gamma}^2+4\tilde{\Lambda}\tilde{\mu}^2)Y_c^4-\nonumber \\
&&(4\tilde{\gamma}^2y_c^4\tilde{\lambda}+(-16\tilde{\gamma}^2\tilde{\mu}^2+24\tilde{\lambda} \tilde{m}^2)y_c^2+8\tilde{m}^2\tilde{\mu}^2)Y_c^2+4y_c^6\tilde{\lambda}^2-4y_c^4\tilde{\mu}^2\tilde{\lambda}\nonumber \\
&&+8y_c^2\tilde{\mu}^4\Big)D_1^{-1}-4\epsilon^2y_c^2-4\epsilon\Big)D_2^{-1},
\end{eqnarray}
\begin{eqnarray}\label{27}
&&M_{14}=-\frac{\sqrt{6}f_{,Y_c}(\epsilon y_c^2-1)^2}{12\epsilon^2y_c^2-2\epsilon y_c^2+2}\nonumber \\
&&=-8\sqrt{6}Y_cy_c(Y_c^4 \tilde{\gamma}^2\tilde{\Lambda}-2\tilde{\Lambda}(\tilde{\lambda} y_c^2-\tilde{\mu}^2)Y_c^2+\tilde{\gamma}^2y_c^4\tilde{\lambda}
-2 \tilde{m}^2 y_c^2\tilde{\lambda}\nonumber \\
&&+2 \tilde{m}^2\tilde{\mu}^2)(\epsilon y_c^2-1)^2(D_1D_2)^{-1},
\end{eqnarray}
\begin{eqnarray}\label{28}
&&M_{52}=\frac{\sqrt{6}}{2}(\epsilon y_c^2-1)F_{,y_c}\nonumber \\
&&=4\sqrt{6}Y_cy_c(Y_c^4 \tilde{\gamma}^2\tilde{\Lambda}-2\tilde{\Lambda}(\tilde{\lambda} y_c^2-\tilde{\mu}^2)Y_c^2+\tilde{\gamma}^2 y_c^4\tilde{\lambda}-2\tilde{m}^2y_c^2\tilde{\lambda}\nonumber \\
&&+2\tilde{m}^2\tilde{\mu}^2)(\epsilon y_c^2-1)D_1^{-1},
\end{eqnarray}
and
\begin{eqnarray}\label{29}
&&M_{54}=\frac{\sqrt{6}}{2}(\epsilon y_c^2-1)F_{,Y_c}\nonumber \\
&&=\sqrt{6}\Big(-4\tilde{\gamma}^2y_c^6\tilde{\lambda}+((-8\tilde{\gamma}^4+12\tilde{\Lambda}\tilde{\lambda})Y_c^2+8\tilde{\gamma}^2\tilde{\mu}^2+4\tilde{\lambda} \tilde{m}^2)y_c^4+\nonumber \\ &&(4Y_c^4\tilde{\gamma}^2\tilde{\Lambda}+(16\tilde{\gamma}^2\tilde{m}^2-24\tilde{\Lambda}\tilde{\mu}^2)Y_c^2-8\tilde{m}^2\tilde{\mu}^2)y_c^2-4Y_c^6\tilde{\Lambda}^2-\nonumber \\
&&4Y_c^4\tilde{m}^2\tilde{\Lambda} -8Y_c^2\tilde{m}^4\Big)(\epsilon y_c^2-1)(2D_3)^{-1},
\end{eqnarray}
where
\begin{equation}\label{30}
D_1:=(\tilde{\Lambda} Y_c^4+2(\tilde{m}^2-\tilde{\gamma}^2y_c^2)Y_c^2+\tilde{\lambda} y_c^4-2\tilde{\mu}^2 y_c^2)^2,
\end{equation}
\begin{equation}\label{31}
D_2:=12\epsilon^2 y_c^2-2\epsilon y_c^2+2,
\end{equation}
and
\begin{equation}\label{31}
D_3:=(\tilde{\lambda} y_c^4-2(\tilde{\mu}^2+\tilde{\gamma}^2Y_c^2)y_c^2+\tilde{\Lambda} Y_c^4+2\tilde{m}^2 Y_c^2)^2.
\end{equation}
We note that $Y_c=\tilde{\psi}_{dS}$ and $y_c=\tilde{\phi}_{dS}$.  The characteristic polynomial of (\ref{24}) is given by
\begin{equation}\label{32}
\mathcal{P}(z)=z^5+\mathcal{P}_4z^4+\mathcal{P}_3z^3+\mathcal{P}_2z^2+\mathcal{P}_1(z)z+\mathcal{P}_0,
\end{equation}
where
\begin{eqnarray}\label{33}
&&\mathcal{P}_4=-M_{11}+\frac{9}{2},\nonumber \\
&&\mathcal{P}_3=-\sqrt{6}M_{12}-\frac{9}{2}M_{11}-\sqrt{6}M_{54}+\frac{9}{2},\nonumber \\
&&\mathcal{P}_2=-\frac{3\sqrt{6}}{2}M_{54}+\sqrt{6}M_{54}M_{11}-\frac{9}{2}M_{11}-\frac{9\sqrt{6}}{2}M_{12},\nonumber \\
&&\mathcal{P}_1=-6M_{52}M_{14}+\frac{3\sqrt{6}}{2}M_{54}M_{11}+6M_{54}M_{12}-\frac{9\sqrt{6}}{2}M_{12},\nonumber \\
&&\mathcal{P}_0=9M_{54}M_{12}-9M_{52}M_{14}.
\end{eqnarray}
Based on Routhâ€“Hurwitz stability criterion, we obtain the necessary and sufficient conditions for the stability at the fixed point
\begin{eqnarray}\label{34}
&&\mathcal{P}_3\mathcal{P}_4-\mathcal{P}_2>0,\nonumber \\
&&\mathcal{P}_3\mathcal{P}_2\mathcal{P}_4-\mathcal{P}_1\mathcal{P}_4^2+\mathcal{P}_0\mathcal{P}_4-\mathcal{P}_2^2>0,\nonumber \\
&&\mathcal{P}_3\mathcal{P}_1\mathcal{P}_2\mathcal{P}_4-\mathcal{P}_3^2\mathcal{P}_0\mathcal{P}_4-\mathcal{P}_1^2\mathcal{P}_4^2
+\mathcal{P}_3\mathcal{P}_0\mathcal{P}_2+2\mathcal{P}_0\mathcal{P}_1\mathcal{P}_4\nonumber\\
&&-\mathcal{P}_1\mathcal{P}_2^2-\mathcal{P}_0^2>0,\nonumber \\
&&\mathcal{P}_4>0,\nonumber \\
&&\mathcal{P}_0>0.
\end{eqnarray}
If (\ref{34}) holds,  all the eigenvalues of (\ref{24}) are negative and $\{y_c,Y_c\}$ is an attractor  point.

So the constraints (\ref{18}), and (\ref{34}) on the parameters $\{\tilde{\mu}, \tilde{m}, \tilde{\gamma},\tilde{\lambda},\tilde{\Lambda}\}$ are the conditions to have a stable de Sitter fixed point.
These are very complicated set of inequalities, and finding their analytical solutions, if not impossible, is very difficult.  But to confirm that such solutions exist, one can find simple numerical solutions satisfying these inequalities.  For examples $\{\tilde{\Lambda} = 1.9999992\times 10^5, \epsilon = \frac{1}{8}, \tilde{\gamma} = 10^6, \tilde{\lambda} = 5000000, \tilde{m} = 1, \tilde{\mu} = 1\}$,  and $\{\tilde{\Lambda}=0.5,\,\, \epsilon=1,\,\, \tilde{\gamma}=1,\,\, \tilde{\lambda}=0,\,\, \tilde{m}=10,\,\, \tilde{\mu}=10\}$  and so on satisfy these constraints.

 \section{Conclusion and discussions}

 To explain the rise of dark energy in a matter dominated era, we considered a Lorentz invariant hybrid cosmological model with a $Z_2$ symmetry. The first quintessence is nonminimally coupled to the Ricci scalar. This coupling relates the matter density to the quintessence effective potential, such that the initial $Z_2$ symmetry breaks by matter dilution. This gives rise to the quintessence evolution from its zero initial expectation value, and for this, in contrast to \cite{khou,sad1,sad2} there is no need to additional conformal coupling. The rolling of the second scalar field, $\psi$, is triggered by the first one which by crossing a critical value, made the $\psi's$ squared effective mass negative. In its turn, the second $Z_2$ symmetry breaking leads to a stable late time attractor solution which is not allowed in single scalar field models \cite{sad1}. Unlike the symmetron model the acceleration is not transient, and eventually the Universe experiences a de Sitter expansion. The stability of the de Sitter point was discussed. The stability depends on the satisfaction of some very complicated conditions (\ref{34}).
\subsection{An illustrative example}
\begin{figure}[H]
\centering\epsfig{file=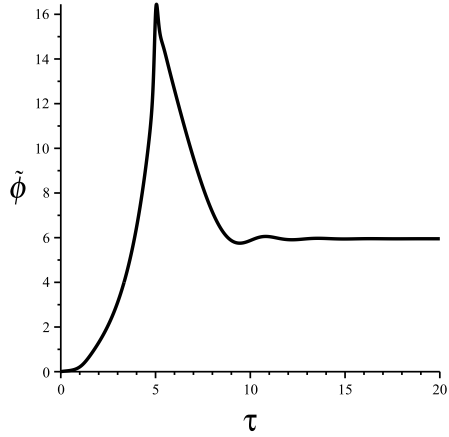,width=6cm}\caption{ Evolution of $\tilde{\phi}$ from $\tau=0$ in the matter dominated era, with initial conditions (IC1), (IC2), and  $\{\epsilon=1,\,\, \tilde{\lambda}=0.005,\,\, \tilde{m}=4,\,\, \tilde{\gamma}=1,\,\, \tilde{\mu}=4,\,\, \tilde{\Lambda}=0.5\}$.} \label{fig1}
\end{figure}

Now to show that how the model works, let us illustrate our results with a simple numerical example. To draw our diagrams, we use the equations (\ref{10}), (\ref{13}), (\ref{14}), and the Friedmann equation
\begin{eqnarray}\label{35}
&&-2\left((6\epsilon^2-\epsilon)\tilde{\phi}^2+1\right)\frac{d\tilde{H}}{d\tau}=(1-2\epsilon)\left(\frac{d\tilde{\phi}}{d\tau}\right)^2+\left(\frac{d\tilde{\psi}}{d\tau}\right)^2
\nonumber \\ &&+8\epsilon \tilde{H} \tilde{\phi} \frac{d\tilde{\phi}}{d\tau}-2\epsilon \mu^2\tilde{\phi}+2\epsilon\lambda\tilde{\phi}^4-2\epsilon\gamma \tilde{\phi}^2 \tilde{\psi}^2+24\epsilon^2\tilde{\phi}^2\tilde{H}^2+\tilde{\rho}^{(m)},
\end{eqnarray}
derived from (\ref{5}). We choose the parameter of the system as $\{\epsilon=1,\,\, \tilde{\lambda}=0.005,\,\, \tilde{m}=4,\,\, \tilde{\gamma}=1,\,\, \tilde{\mu}=4,\,\, \tilde{\Lambda}=0.5\}$. Inserting these  parameters in (\ref{17}), we obtain $\{\tilde{H}_{dS}^2 = 4.55, \tilde{\phi}_{dS}^2 = 35.41, \tilde{\psi}_{dS}^2 = 38.83\}$. To depict the behavior of the system, we must set initial conditions at the time after which the quintessence, $\phi$, becomes tachyonic. We take this time as $\tau_1=0$. i.e.  $\tilde{\rho}^{(m)}(\tau\geq 0)\leq \frac{\tilde{\mu}^2}{\epsilon}$. Before this time, i.e. when the matter density satisifies $\tilde{ \rho}^{(m)}>16$, the scalar fields does not contribute in the Universe evolution. We choose the following initial conditions in the matter dominated era (we neglect the radiation contribution):
$IC1:=\{H(0)=2.311,\,\, \tilde{\rho}^{(m)}(0)=16 ,\,\, \tilde{\psi}(0)=0,\,\,  \frac{d\tilde{\psi}}{d\tau }(0)=0,\,\, \tilde{\phi}(0)=0,\,\,  \frac{d\tilde{\phi}}{d\tau }(0)=0.2 \}$. In fig.(\ref{fig1}), the evolution of $\tilde{\phi}$ from $\tau=\tau_1=0$ is plotted in terms of the dimensionless time $\tau$.
This figure shows that $\tilde{\phi}$ increases from its negligible initial value and finally settle at $\tilde{\phi}_{dS}=5.95$.

\begin{figure}[H]
\centering\epsfig{file=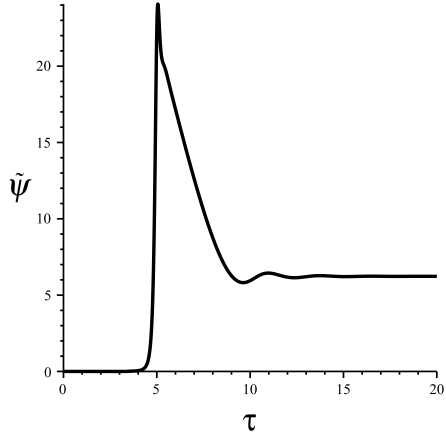,width=6cm,angle=0}
\caption{ Evolution of $\tilde{\psi}$ from $\tau=0$ in the matter dominated era, with initial conditions (IC1) and (IC2) and  $\{\epsilon=1,\,\, \tilde{\lambda}=0.005,\,\, \tilde{m}=4,\,\, \tilde{\gamma}=1,\,\, \tilde{\mu}=4,\,\, \tilde{\Lambda}=0.5\}$}\label{fig2}
\end{figure}

Evolution of $\tilde{\psi}$ begins from
$\tau=3.338$, the time at which $\tilde{\phi}=\sqrt{\frac{\tilde{m}^2}{\tilde{\gamma}}}$. At this point we have $\{\tilde{\phi}(2.338)=4,\,\,  \frac{d\tilde{\phi}}{d\tau }(3.338)=2.968,\,\,\tilde{\rho}^{(m)}(3.338)=0.001 \}$. The initial condition for $\psi$ is set as $IC2:=\{\tilde{\psi}(3.338)=0,\,\,  \frac{d\tilde{\psi}}{d\tau }(3.338)=0.1\}$. The evolution of $\tilde{\psi}$ is demonstrated  in fig.(\ref{fig2}).
Like $\phi$, $\psi$ settles down eventually at the fixed point.

\begin{figure}[H]
\begin{subfigure}[h]{0.5\textwidth}
\centering
\includegraphics[width=1.1\textwidth,trim=3mm 4mm 25mm 2mm, clip=true]{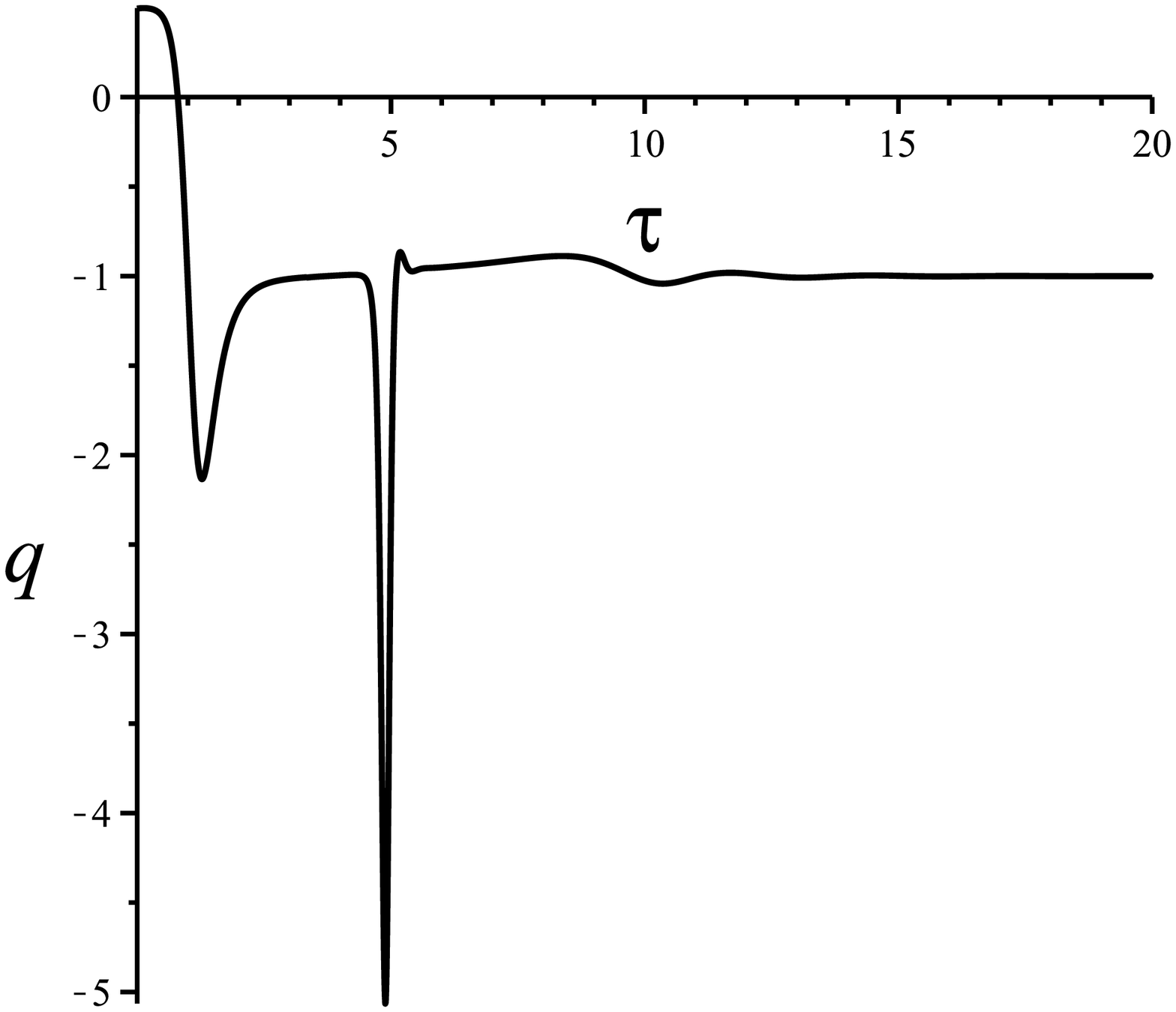}
\caption{}\label{fig31}
\end{subfigure}
\begin{subfigure}[h]{0.5\textwidth}
\centering
\sbox0{\includegraphics[width=0.5\textwidth,trim=3mm 4mm 25mm 2mm, clip=true]{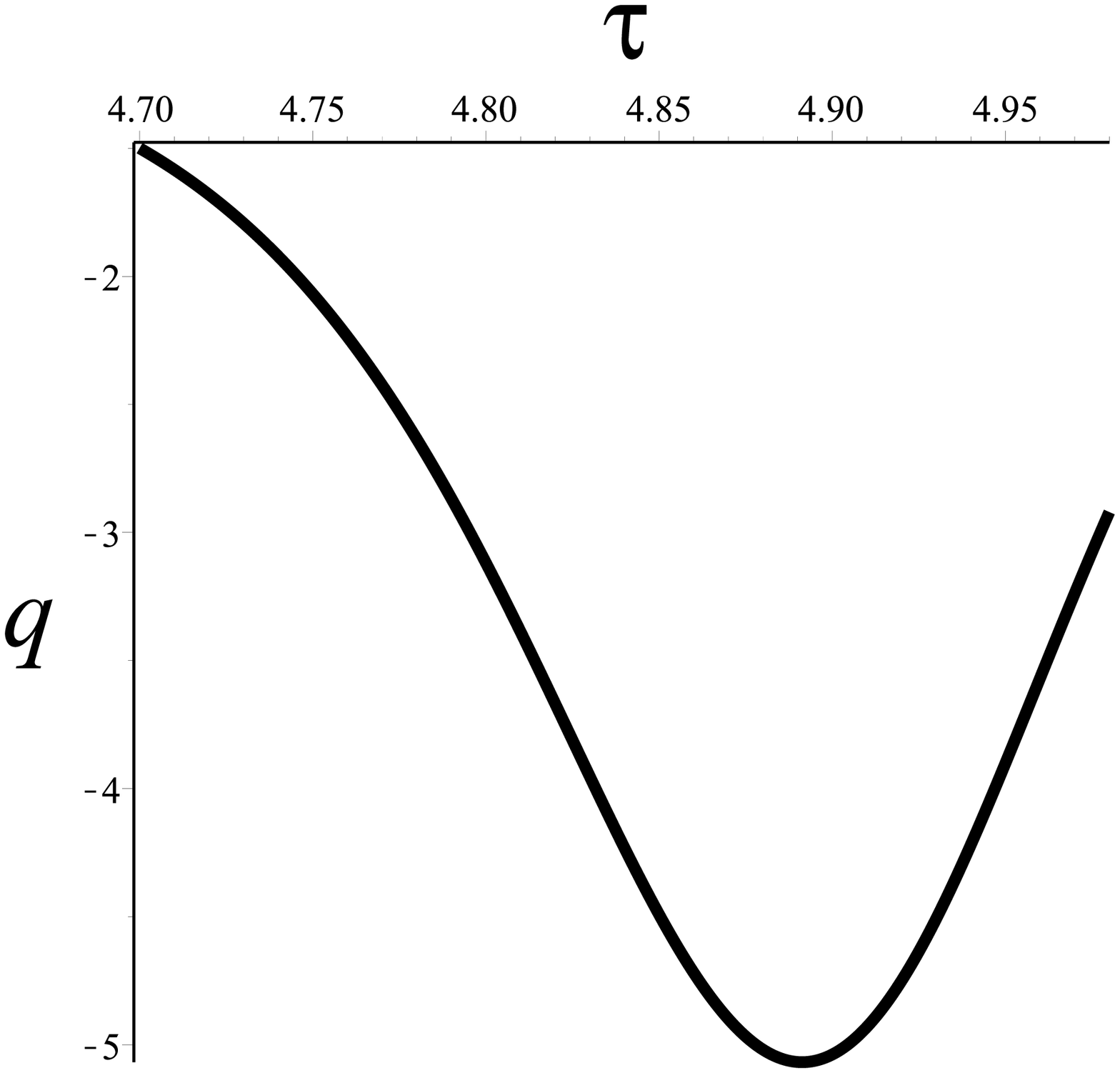}}
\includegraphics[height=\ht0,keepaspectratio]{2.eps}
\caption{}\label{fig32}
\end{subfigure}
\caption{(a) The deceleration parameter in terms of dimensionless time $\tau=tH_0$, from $\tau=0$ in the matter dominated era, with initial conditions (IC1) and (IC2) and $\{\epsilon=1,\,\, \tilde{\lambda}=0.005,\,\, \tilde{m}=4,\,\, \tilde{\gamma}=1,\,\, \tilde{\mu}=4,\,\, \tilde{\Lambda}=0.5\}$, (b) Behavior of $q$ near the peak.}
\end{figure}

\begin{figure}[H]
\centering
\begin{subfigure}[b]{0.5\linewidth}
\includegraphics[width=\linewidth]{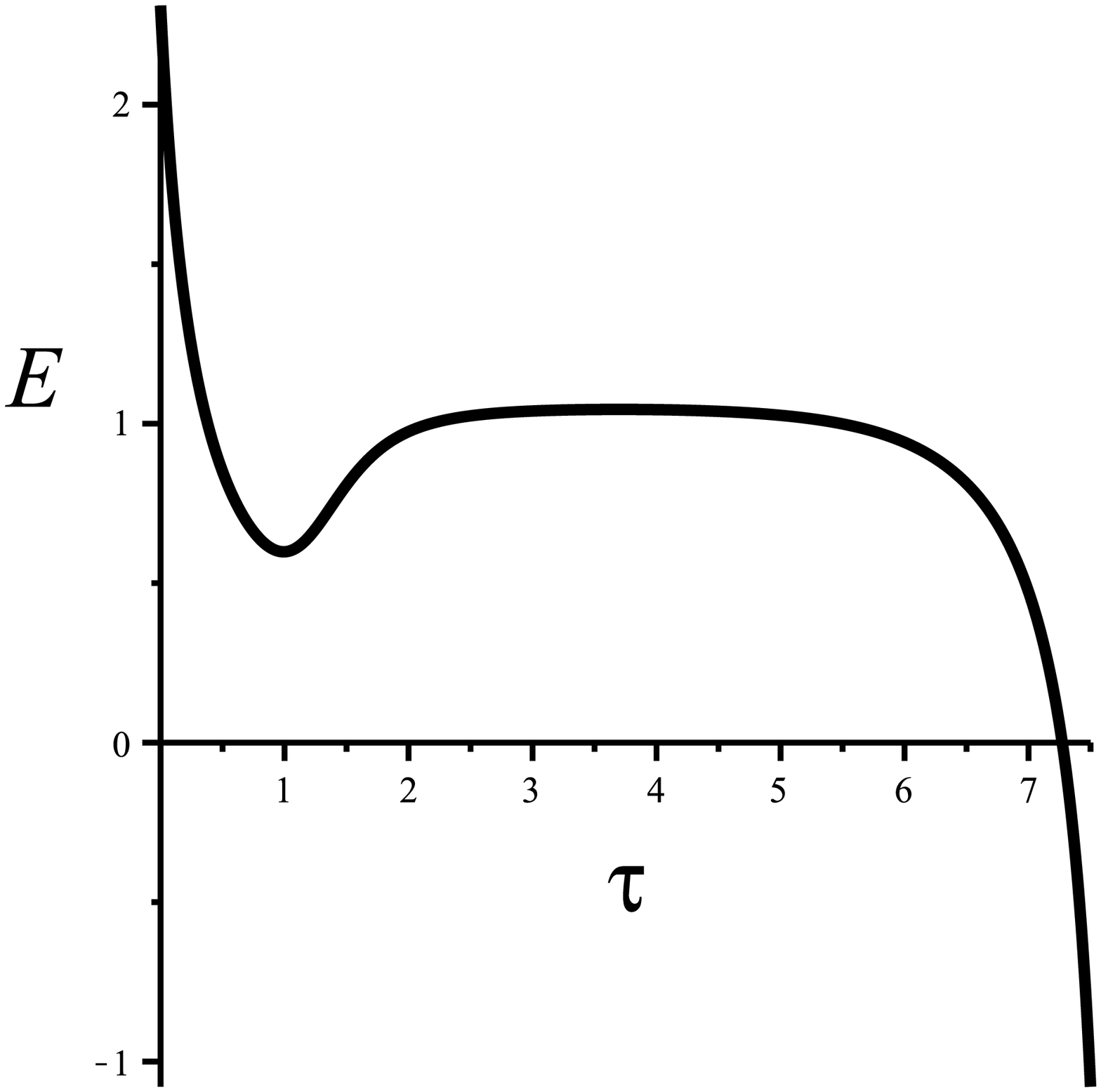}
\end{subfigure}
\begin{subfigure}[b]{0.5\linewidth}
\includegraphics[width=\linewidth]{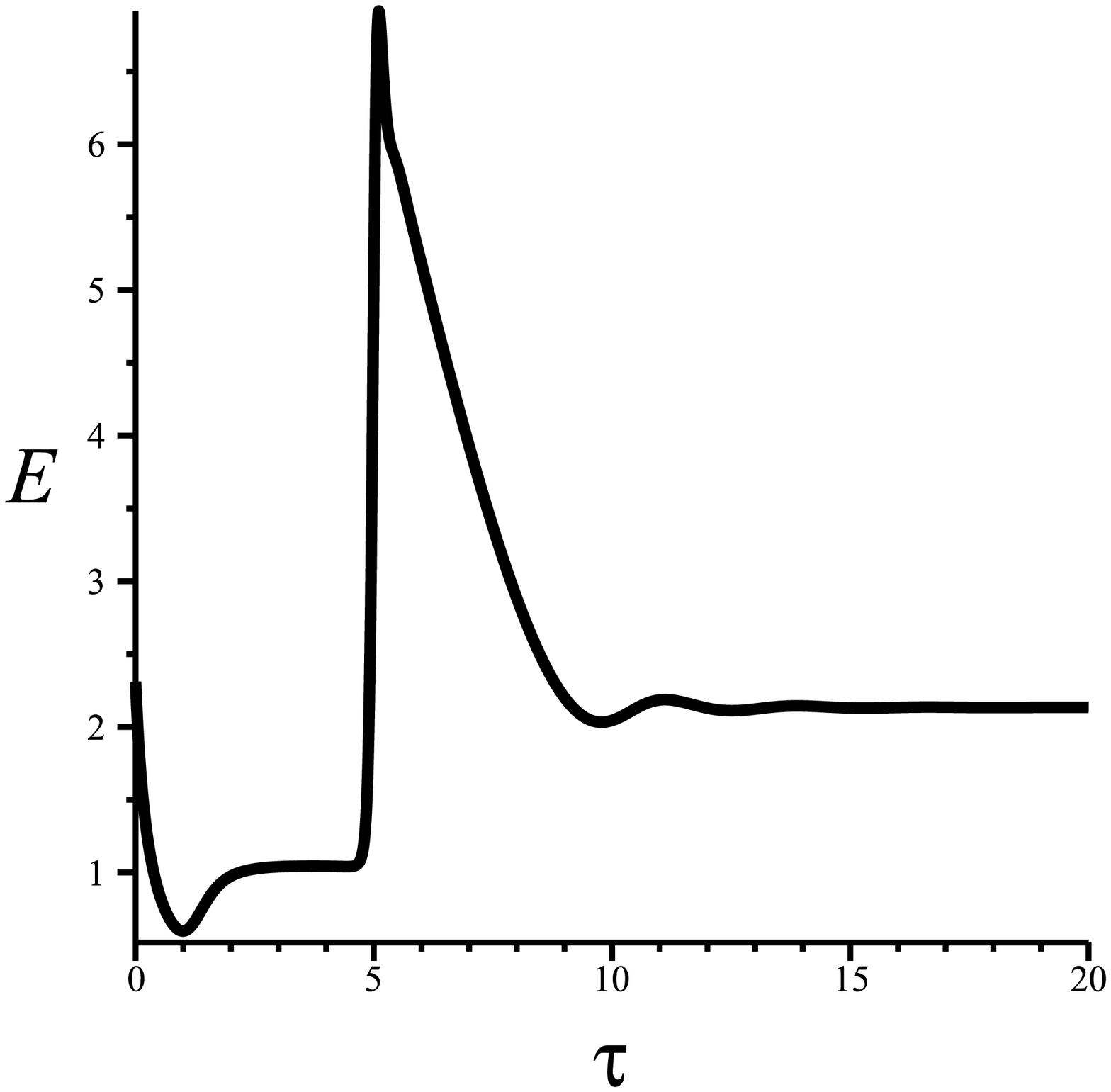}
\end{subfigure}
\caption{The dimensionless Hubble parameter $\frac{H}{H_0}$ labeled by $E$, in terms of $\tau$, with initial conditions (IC1) and (IC2) and $\{\epsilon=1,\,\, \tilde{\lambda}=0.005,\,\, \tilde{m}=4,\,\, \tilde{\gamma}=1,\,\, \tilde{\mu}=4,\,\, \tilde{\Lambda}=0.5\}$, in the absence of the second scalar field (up), and when the second scalar field is considered (down).}
\label{fig4}
\end{figure}

\begin{figure}[H]
\centering\epsfig{file=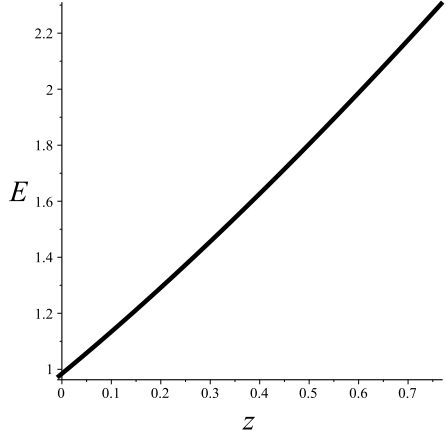,width=6cm,angle=0}
\caption{ The dimensionless Hubble parameter $\frac{H}{H_0}$ labeled by $E$, in terms of the redshift with initial conditions (IC1) and (IC2) $\{\epsilon=1,\,\, \tilde{\lambda}=0.005,\,\, \tilde{m}=4,\,\, \tilde{\gamma}=1,\,\, \tilde{\mu}=4,\,\, \tilde{\Lambda}=0.5\}$,  from $\tau=0$ until the present time $z=0$}\label{fig33}
\end{figure}

The deceleration parameter $q$ defined by  $q=-1-\frac{\dot{H}}{H^2}$ , is depicted in fig.(\ref{fig31}). This figure shows that initially where $\rho^{(m)}$ is dominant, the deceleration parameter is given by $q=\frac{1}{2}(1+3w^{(m)})=\frac{1}{2}$. Then, the Universe experiences a positive acceleration and finally stays at a de Sitter point $\dot{H}=0$. At $\tau=0.9$, we have $q=-0.5$ and $\frac{ \rho^{(m)}}{3M_P^2H_0^2}=0.3$ which is similar to a situation like our present time.  An analytical analysis of the behavior of $q$ in our nonminimally coupled hybrid model is very difficult. In minimal ($\epsilon=0$) screening models like the symmetron model, $\rho_d$ and $P_d$, appearing in the Friedmann eqs.(\ref{4},\ref{5}), do not contain additional expressions in terms of the Hubble parameter (see (\ref{6}),(\ref{7})) , therefore the dynamical behavior of the system is less complicated.  For example, in contrast to minimal models, we may have situations with $q<-1$ which occurs when the Universe super accelerates, i.e $\dot{H}>0$. This is a commune aspect of models with non minimal coupling to the Ricci scalar\cite{far}. So the time derivative of our functions like the Hubble parameter may become zero at some points, and some peaks appear in our diagrams as shown in fig.(\ref{fig32}).
The Hubble parameter is demonstrated in fig.(\ref{fig4}), showing different behaviors.  It decreases in the matter dominated era, then after a Hubble time it increases giving rise to a super acceleration phase, after that it becomes almost a constant and after some fluctuations it will eventually reach a steady value.
Another interesting point is the r\^{o}le of the second scalar field to avoid the turnarounds and singularities. For the above numerical example, if we ignored $\psi$, the Universe would experience a turnaround at $\tau=7.28$ see fig.(\ref{fig4}). This is avoided by activation of $\psi$ at $\tau=3.338$.
Note that in our example, the super acceleration ($\dot{H}>0$) has not happened in the past see figures (\ref{fig4},\ref{fig33}). We have a super acceleration period after $\tau\simeq 1$. If this happened after matter domination, it could be observed or ruled out by BAO or Ly-$\alpha$ forest, or even 21cm line observations \cite{ref}.

\subsection{Parameters estimation}
To specify the parameters, we note that if the acceleration expansion has started from $z=z_*$, our model requires that the symmetry be broken before that time. So we must take $\rho^{m}(z_*)<\rho_{cr}$. In addition, one can consider some additional cosmological assumptions, e.g. if we demand that BBN (Big Bang nucleosynthesis) precise predictions are not affected by $\phi$, we may assume that the symmetry was stored at that time, such that $\phi$ and $\psi$ were frozen at $\phi=\psi=0$. So we must take the critical density less than the matter density at BBN. Using $\rho^{m}(z)=\rho^{m}(z=0)(1+z)^3$, where $z$ is the redshift  we obtain $\rho_{cr.}<(1+z_{BBN})^{3}\rho^{(m)}(z=0)$.  Combining these with the Planck 2015 data : $\rho^{(m)}(z=0)\simeq 4.4\times 10^{-10} eV^4(=4.12\times 10^{-31} gr/cm^3)$, we obtain

\begin{equation}\label{36}
(1+z_*)^3\times4.12\times 10^{-31} gr/cm^3<\rho_{cr}<(1+z_{BBN})^{3}\times4.12\times 10^{-31} gr/cm^3
\end{equation}
Taking  $H_0=67.8 km s^{-1}Mpc^{-1}$ \cite{pla}, (\ref{36}) gives
\begin{equation}\label{37}
0.924(1+z_*)^3<\frac{\tilde{\mu}^2}{\epsilon}<0.924(1+z_{BBN})^{3}
\end{equation}
But $z_{BBN}\gg z_*$, so it seems that in principle one can choose $\mu$ and $\epsilon$ such that (\ref{37}) be satisfied. For $z^*\simeq 0.6$ and $z_{BBN}\sim 10^8$,  we have $3.7<\frac{\tilde{\mu}^2}{\epsilon}<9\times 10^{23}$ and $1.7\times 10^{-30} gr/cm^3<\rho_{cr.}<4.21\times 10^{-7} gr/cm^3$. The upper huge value for $\frac{\tilde{\mu}^2}{\epsilon}$ is related to the fact that the redshift at BBN is much greater than $z^*$. To reduce the upper bound, one may utilize other cosmological data or even local gravitational data as follows:
 In scalar field models, if there is a coupling between the scalar field and the matter, a new force arises in the matter section, dubbed fifth force. This force is mediated by the scalar fields. In the vicinity of a matter source (like the Sun or the Earth and ..)  this coupling is coming from the term $\epsilon R\phi^2\simeq \frac{\epsilon}{M_P^2}{\rho^{(m)}}\phi^2$ (see (\ref{1})).  The fifth force between matter sources is governed by the equation \cite{khs,inter}
\begin{equation}\label{38}
\Box \phi -V_{,\phi}=\frac{\epsilon}{M_P^2}\rho^{(m)}\phi.
\end{equation}
Note that the other field, i.e. $\psi$, is not coupled to the curvature, so it does not mediate any force in the matter sector
\begin{equation}\label{39}
\Box \psi -V_{,\psi}=0.
\end{equation}
In the dense regions where the quintessence is completely screened and we have $<\phi>=0$. Therefore the fifth force vanishes. If we assume that the nonrealistic matter density in the solar system or in our galaxy is greater than the critical density $\rho_{cr.}=\frac{\tilde{\mu}^2}{\epsilon}M_p^2H_0^2$, the fifth force is ignorable. Indeed if in a region the matter density exceeds the critical density, the symmetry is restored deep inside, and the solutions of the field equations ((\ref{38}),(\ref{39})) become $\phi=\psi=0$. As there is no strong evidence for the fifth force in local gravitational tests, it is desirable to find conditions that the quintessence while drives the cosmic acceleration, is screened in our local environments. The dark matter density in our solar system in the vicinity of the Sun is estimated to be $\rho^{(dm)}_{Sun}\simeq 7.13\times 10^{-25}gr/cm^3$ \cite{sun}, while in the vicinity of the Earth, Mars and Saturn the estimation is $\rho^{(dm)}_{E}<1.4\times 10^{-19}gr/cm^3$, $\rho^{(dm)}_{M}<1.4\times 10^{-20}gr/cm^3$, and $\rho^{(dm)}_{S}<1.1\times 10^{-20}gr/cm^3$ \cite{planets}. In the Milky Way, the dark matter is not uniformly distributed, and for a spherical dark matter halo profile may be in the domain $3.56-9.98 \times 10^{-25}gr/cm^3$ \cite{milk}.  So to have screening in these areas it is necessary that these densities be larger than the critical density $\rho_{cr.}<10^{-25}gr/cm^3$. This gives $\frac{\tilde{\mu}^2}{\epsilon}<2.2\times 10^5$,
which tightens the domain obtained in (\ref{36}).

At the end let us note that the gravitational effects of the nonminimal coupling to the curvature in a general hybrid model have also been carried out in the context of the parameterized post-Newtonian approximation (PPN)in \cite{jar}. There was shown that in the lowest order approximation, $V(\phi, \psi)=0$ and $V_{,\phi}(\phi,\psi)=V_{,\psi}(\phi,\psi)=0$ hold, which for the potential (\ref{11}) gives : $\phi=\psi=0$.   This implies that the non minimal coupling does not affect the parameterized post-Newtonian parameters in gravitational test\cite{jar} which is consistent with the screening effect used in our article.

\end{document}